\documentclass[prd,twocolumn,tightenlines,showpacs,nofootinbib]{revtex4-1} 

\usepackage{amsfonts}
\usepackage{dcolumn}
\usepackage{bm}
\usepackage{amsmath}
\usepackage{nicefrac}
\usepackage{amsthm}
\usepackage{amssymb}
\usepackage{graphicx}
\usepackage{color}

\newcommand{\appropto}{\mathrel{\vcenter{
  \offinterlineskip\halign{\hfil$##$\cr
    \propto\cr\noalign{\kern2pt}\sim\cr\noalign{\kern-2pt}}}}}

\begin{document}

\title{Nuclear spin-dependent interactions: Searches for WIMP, Axion and Topological Defect Dark Matter, and Tests of Fundamental Symmetries}

\date{\today}

\author{Y.~V.~Stadnik} \email[]{y.stadnik@unsw.edu.au}
\author{V.~V.~Flambaum} 
\affiliation{School of Physics, University of New South Wales, Sydney 2052,
Australia}

\begin{abstract}

We calculate the proton and neutron spin contributions for nuclei using semi-empirical methods, as well as a novel hybrid \emph{ab initio}/semi-empirical method, for interpretation of experimental data. We demonstrate that core-polarisation corrections to \emph{ab initio} nuclear shell model calculations generally reduce discrepancies in proton and neutron spin expectation values from different calculations. We derive constraints on the spin-dependent P,T-violating interaction of a bound proton with nucleons, which for certain ranges of exchanged pseudoscalar boson masses improve on the most stringent laboratory limits by several orders of magnitude. We derive a limit on the CPT and Lorentz-invariance-violating parameter $|\tilde{b}_{\perp}^p| < 7.6 \times 10^{-33}$ GeV, which improves on the most stringent existing limit by a factor of 8, and demonstrate sensitivities to the parameters $\tilde{d}_{\perp}^p$ and $\tilde{g}_{ D\perp}^p$ at the level $\sim 10^{-29} - 10^{-28}$ GeV, which is a one order of magnitude improvement compared to the corresponding existing sensitivities. We extend previous analysis of nuclear anapole moment data for Cs to obtain new limits on several other CPT and Lorentz-invariance-violating parameters: $\left|b_0^p \right| < 7 \times 10^{-8}$ GeV, $\left|d_{00}^p \right| < 8 \times 10^{-8}$, $\left|b_0^n \right| < 3 \times 10^{-7}$ GeV and $\left|d_{00}^n \right| < 3 \times 10^{-7}$. 

\end{abstract}

\maketitle

\section{Introduction}
The violation of the fundamental symmetries of nature is an active area of research. Atomic and molecular experiments, which probe $\mathcal{P}$-odd and $\mathcal{P}$,$\mathcal{T}$-odd interactions, provide very sensitive tests of the Standard Model (SM) and physics beyond the SM \cite{Khriplovich1991_PNC-Book,Ginges_PhysRep2004,Roberts2015-Review(PNC)}. Measurements and calculations of the Cs $6s$-$7s$ parity nonconserving (PNC) amplitude stand as the most precise atomic test of the SM electroweak theory to date, see e.g.~\cite{Bouchiat1982Cs,Wood1997_Cs-PNC,Dzuba1989Cs,Blundell92Cs,Kozlov01Cs,Safronova02Cs,Ginges02Cs,Roberts2012_Cs-pnc}. Experimental searches for nuclear anapole moments are ongoing in Fr \cite{Aubin2013_Fr-AM-project}, Yb \cite{Tsigutkin2009_Yb-PNC,Tsigutkin2010_Yb-pnc} and BaF \cite{Demille2008,DeMille2014-BaF(Zeeman)}. At present, Hg provides the most precise limits on the electric dipole moment (EDM) of the proton, quark chromo-EDM and $\mathcal{P}$,$\mathcal{T}$-odd nuclear forces, as well as the most precise limits on the neutron EDM and quantum chromodynamics (QCD) $\theta$ term from atomic or molecular experiments \cite{Griffith2009improved,Swallows2013}, while ThO provides the most precise limit on the electron EDM \cite{Baron2014}. Most recently, it was suggested that EDM measurements in molecules with $\mathcal{P}$,$\mathcal{T}$-odd nuclear magnetic quadrupole moments may lead to improved limits on the strength of $\mathcal{P}$,$\mathcal{T}$-odd nuclear forces, proton, neutron and quark EDMs, quark chromo-EDM and the QCD $\theta$ term \cite{Flambaum2014-MQM}. 

Field theories, which are constructed from the principles of locality, spin-statistics and Lorentz invariance, conserve the combined $\mathcal{CPT}$ symmetry. The violation of one or more of these three principles, presumably from some form of ultra-short distance scale physics, opens the door for the possibility of $\mathcal{CPT}$-odd physics. Some of the most stringent limits on $\mathcal{CPT}$-odd and Lorentz-invariance-violating physics come from searches for the coupling $\mathbf{\tilde{b}}\cdot\mathbf{s}$ between a background cosmic field, $\mathbf{\tilde{b}}$, and the spin of an electron, proton, neutron or muon, $\mathbf{s}$ \cite{Berglund1995,Bluhm2000a,Hou2003,Cane2004,Heckel2006,Heckel2008,Bennett2008CPT,Altarev2009,Gemmel2010,Brown2010,Peck2012,Allmendinger2014}. For further details on the broad range of experiments performed and a brief history of the improvements in these limits, we refer the reader to the reviews of \cite{Kostelecky1999,Kostelecky2011data} and the references therein. 

Other very important unanswered questions in fundamental physics are the strong $\mathcal{CP}$ problem, namely the puzzling observation that QCD does not appear to violate the combined charge-parity ($\mathcal{CP}$) symmetry, see e.g.~\cite{Weinberg1976,Weinberg1978,Peccei1977a,Peccei1977b,Wilczek1978,Moody1984},
and dark matter and dark energy, see e.g.~\cite{Bertone2005,Spergel2007,Agnese2013,Riess1998,Perlmutter1999}. A particularly elegant solution to the strong {$\mathcal{CP}$} problem invokes the introduction of a pseudoscalar particle known as the axion~\cite{Peccei1977a,Peccei1977b} (see also \cite{Kim1979,Shifman1980a,Zhitnitsky1980,Dine1981a}). It has been noted that the axion may also be a promising cold dark matter candidate. Thus axions, if detected, could resolve both the dark matter and strong {$\mathcal{CP}$} problems~\cite{Kim2010,Kawasaki2013,Brambilla2014,Baer2014,Stadnik2014_Axions-review}. The decay of supersymmetric axions to produce axions has also been suggested as a possible explanation for dark radiation \cite{Jeong2012,Graf2013a,Graf2013b,Queiroz2014}.

Many tests of the fundamental symmetries of nature and searches for axion, weakly-interacting massive particle (WIMP) and topological defect dark matter involve couplings of the form $\mathbf{X} \cdot \mathbf{s}_N$ between a field or operator $\mathbf{X}$ and the spin angular momentum $\mathbf{s}_N$ of a proton ($N=p$) or neutron ($N=n$), or depend explicitly on the spin angular momenta of the nucleons involved. We point out that in experiments, which measure nuclear spin-dependent (NSD) properties, the contribution of non-valence nucleon spins cannot be neglected, due to polarisation of these nucleons by the valence nucleon(s). Nuclear many-body effects have previously been considered in association with the interpretation of atomic clock experiments \cite{Flambaum2003Nuc,Flambaum2006a,Berengut2011K}, nuclear-sourced EDMs and NSD-PNC interactions mediated via $Z^0$-boson exchange between electrons and the nucleus (see e.g.~\cite{Berengut2011K}), static spin-gravity couplings \cite{Flambaum2009a,Kimball2014_Nuc} and long-range dipole-dipole couplings \cite{Kimball2014_Nuc}.

In the present work, we calculate the proton and neutron spin contributions for a wide range of nuclei, which are of experimental interest in tests of the fundamental symmetries of nature and searches for dark matter, including axions, WIMPs and topological defects, using semi-empirical methods, as well as a novel hybrid \emph{ab initio}/semi-empirical method. We then demonstrate that core-polarisation corrections to \emph{ab initio} nuclear shell model calculations generally reduce discrepancies in proton and neutron spin expectation values from different calculations. As an illustration of the importance of many-body effects in such studies, we revisit the experiments of Refs.~\cite{Gemmel2010,Allmendinger2014}, in which a $^{3}$He/$^{129}$Xe comagnetometer was used to place constraints on the $\mathcal{CPT}$ and Lorentz-invariance-violating parameter $\tilde{b}_{\perp}^n$, which quantifies the interaction strength of a background field with the spin of a neutron. We show that, due to nuclear many-body effects, the $^{3}$He/$^{129}$Xe system is in fact also quite sensitive to proton interaction parameters. By reanalysing the results of Ref.~\cite{Allmendinger2014}, we derive a limit on the parameter $\tilde{b}_{\perp}^p$ that is the world's most stringent by a factor of 8. Likewise, by reanalysing the results of Ref.~\cite{Cane2004}, in which a $^{3}$He/$^{129}$Xe comagnetometer was also used, we demonstrate improved sensitivities to the parameters $\tilde{d}_{\perp}^p$ and $\tilde{g}_{D \perp}^p$ by one order of magnitude. From existing data in Ref.~\cite{Tullney2013}, in which experiments were performed with a $^{3}$He/$^{129}$Xe comagnetometer, we derive constraints on the spin-dependent $\mathcal{P}$,$\mathcal{T}$-violating interaction of a bound proton with nucleons, which for certain ranges of exchanged pseudoscalar boson masses improve on the most stringent laboratory limits by several orders of magnitude. We also extend our previous analysis of nuclear anapole moment data for Cs \cite{Roberts2014} to obtain new limits on several other $\mathcal{CPT}$ and Lorentz-invariance-violating parameters.

\section{Nuclear Theory}
The nuclear magnetic dipole moment $\mu$ can be expressed (in the units of the nuclear magneton $\mu_N = e\hbar / 2 m_N$):
\begin{equation}
\label{mu-defn}
\mu = g_p \left<s_p^z\right> + g_n \left<s_n^z\right> + \left<l_p^z\right> ,
\end{equation}
where $\left<s_p^z\right>$ and $\left<s_n^z\right>$ are the expectation values of the total proton and neutron spin angular momenta, respectively, while $\left<l_p^z\right>$ is the expectation value of the total proton orbital angular momentum. In the present work, we consider nuclei with either one valence proton or one valence neutron (even-even nuclei are spinless due to the nuclear pairing interaction).

We start by considering the contribution of the valence nucleon alone. Assuming all other nucleons in the nucleus are paired (and ignoring polarisation of the nuclear core for now), the spin $I$ and nuclear magnetic dipole moment $\mu$ are due entirely to the total angular momentum of the external nucleon: $\mathbf{I}=\mathbf{j}=\mathbf{l}+\mathbf{s}$. In this case, the nuclear magnetic dipole moment is given by the Schmidt (single-particle approximation) formula
\begin{equation}
\label{Schmidt_mu}
\mu^0 = g_s \left<s_z\right>^0 + g_l \left<l_z\right>^0 ,
\end{equation}
with
\begin{eqnarray}
\label{sz0}
\left<s_z\right>^0 = \left\{ \begin{array}{ll}
\frac{1}{2} & \textrm{if $j=l+\frac{1}{2}$,}\\
-\frac{j}{2(j+1)} & \textrm{if $j=l-\frac{1}{2}$,}
\end{array} \right.
\end{eqnarray}
\begin{eqnarray}
\label{lz0}
\left<l_z\right>^0 = \left\{ \begin{array}{ll}
j - \frac{1}{2} & \textrm{if $j=l+\frac{1}{2}$,}\\
\frac{j(2j+3)}{2(j+1)} & \textrm{if $j=l-\frac{1}{2}$.}
\end{array} \right.
\end{eqnarray}
The gyromagnetic factors are: $g_l = 1, g_s = g_p = 5.586$ for a valence proton and $g_l = 0, g_s = g_n = -3.826$ for a valence neutron. We present the values for $\left<s_z\right>^0$ from Eq.~(\ref{sz0}) (``Schmidt model'') in Tables \ref{tab:odd-proton_isotopes} and \ref{tab:odd-neutron_isotopes}.

  \begin{table}[h!]
    \centering%
    \caption{Expectation values $\left<s_p^z\right>$ and $\left<s_n^z\right>$ for selected odd-proton nuclei. Nuclear spin and parity assignments, and experimental values of $\mu$ were taken from Ref.~\cite{lbnl_nuclear_isotopes}.} 
\begin{tabular}{cccccc}
 & \multicolumn{1}{c}{Schmidt model}   &  \multicolumn{2}{c}{Minimal model}  &  \multicolumn{2}{c}{Preferred model}  \\ 
\cline{2-2} \cline{3-4} \cline{5-6} 
\multicolumn{1}{c}{Nucleus}   & \multicolumn{1}{c}{$\left<s_z\right>^0$}   & \multicolumn{1}{c}{$\left<s_p^z\right>$} & \multicolumn{1}{c}{$\left<s_n^z\right>$}  & \multicolumn{1}{c}{$\left<s_p^z\right>$} & \multicolumn{1}{c}{$\left<s_n^z\right>$} \\ 
\hline
$^{1}$H	&	0.500	&	0.500	&	0.000	&	0.500	&	0.000	\\
$^{7}$Li	&	0.500	&	0.443	&	0.057	&	0.436	&	0.064	\\
$^{19}$F 	&	0.500	&	0.483	&	0.017	&	0.480	&	0.020	\\
$^{23}$Na	&	-0.300	&	-0.078	&	-0.222	&	-0.051	&	-0.249	\\
$^{27}$Al	&	0.500	&	0.378	&	0.122	&	0.363	&	0.137	\\
$^{35}$Cl	&	-0.300	&	-0.226	&	-0.074	&	-0.217	&	-0.083	\\
$^{39}$K	&	-0.300	&	-0.272	&	-0.028	&	-0.268	&	-0.032	\\
$^{41}$K	&	-0.300	&	-0.290	&	-0.010	&	-0.289	&	-0.011	\\
$^{69}$Ga	&	0.500	&	0.311	&	0.189	&	0.289	&	0.211	\\
$^{81}$Br	&	0.500	&	0.338	&	0.162	&	0.319	&	0.181	\\
$^{85}$Rb	&	-0.357	&	-0.305	&	-0.052	&	-0.299	&	-0.058	\\
$^{87}$Rb	&	0.500	&	0.389	&	0.111	&	0.376	&	0.124	\\
$^{93}$Nb	&	0.500	&	0.434	&	0.066	&	0.426	&	0.074	\\
$^{127}$I 	&	0.500	&	0.290	&	0.210	&	0.265	&	0.235	\\
$^{133}$Cs	&	-0.389	&	-0.297	&	-0.092	&	-0.286	&	-0.103	\\
$^{139}$La	&	-0.389	&	-0.276	&	-0.113	&	-0.262	&	-0.127	\\
$^{141}$Pr	&	0.500	&	0.445	&	0.055 	&	0.438	&	0.062	\\
$^{159}$Tb	&	-0.300 	&	-0.099	&	-0.201	&	-0.075	&	-0.225	\\
$^{165}$Ho	&	0.500	&	0.324	&	0.176	&	0.303	&	0.197	\\
$^{169}$Tm&	0.500	&	0.179	&	0.321	&	0.140	&	0.360	\\
$^{203}$Tl	&	0.500	&	0.376	&	0.124	&	0.361	&	0.139	\\
$^{205}$Tl	&	0.500	&	0.377	&	0.123	&	0.363	&	0.137	\\
$^{209}$Bi	&	-0.409	&	-0.251	&	-0.158	&	-0.232	&	-0.177	\\
$^{209}$Fr	&	-0.409	&	-0.268	&	-0.141	&	-0.251	&	-0.158	\\
$^{211}$Fr	&	-0.409	&	-0.263	&	-0.146	&	-0.246	&	-0.164	\\
  \end{tabular}%
    \label{tab:odd-proton_isotopes}%
  \end{table}

  \begin{table}[h!]
    \centering%
    \caption{Expectation values $\left<s_n^z\right>$ and $\left<s_p^z\right>$ for selected odd-neutron nuclei. Nuclear spin and parity assignments, and experimental values of $\mu$ were taken from Ref.~\cite{lbnl_nuclear_isotopes}.} 
\begin{tabular}{cccccc}
 & \multicolumn{1}{c}{Schmidt model}   &  \multicolumn{2}{c}{Minimal model}  &  \multicolumn{2}{c}{Preferred model}  \\ 
\cline{2-2} \cline{3-4} \cline{5-6} 
\multicolumn{1}{c}{Nucleus}   & \multicolumn{1}{c}{$\left<s_z\right>^0$}   & \multicolumn{1}{c}{$\left<s_n^z\right>$} & \multicolumn{1}{c}{$\left<s_p^z\right>$}  & \multicolumn{1}{c}{$\left<s_n^z\right>$} & \multicolumn{1}{c}{$\left<s_p^z\right>$} \\  
\hline
$^{3}$He	&	0.500	&	0.500	&	0.000	&	0.500	&	0.000	\\
$^{9}$Be	&	0.500	&	0.422	&	0.078	&	0.413	&	0.087	\\
$^{13}$C	&	-0.167	&	-0.174	&	0.007	&	-0.174	&	0.008	\\
$^{21}$Ne	&	-0.300	&	-0.108	&	-0.192	&	-0.085	&	-0.215	\\
$^{29}$Si	&	0.500	&	0.356	&	0.144	&	0.339	&	0.161	\\
$^{39}$Ar	&	0.500	&	0.43	&	0.07	&	0.43	&	0.07	\\
$^{73}$Ge 	&	0.500	&	0.390	&	0.110	&	0.377	&	0.123	\\
$^{87}$Sr	&	0.500	&	0.413	&	0.087	&	0.403	&	0.097	\\
$^{91}$Zr	&	0.500	&	0.435	&	0.065	&	0.428	&	0.072	\\
$^{125}$Te	&	0.500	&	0.391	&	0.109	&	0.378	&	0.122	\\
$^{129}$Xe	&	0.500	&	0.379	&	0.121	&	0.365	&	0.135	\\
$^{131}$Xe 	&	-0.300	&	-0.252	&	-0.048	&	-0.246	&	-0.054	\\
$^{135}$Ba	&	-0.300	&	-0.267	&	-0.033	&	-0.263	&	-0.037	\\
$^{137}$Ba	&	-0.300	&	-0.278	&	-0.022	&	-0.275	&	-0.025	\\
$^{171}$Yb	&	-0.167	&	-0.151	&	-0.015	&	-0.150	&	-0.017	\\
$^{173}$Yb	&	-0.357	&	-0.140	&	-0.217	&	-0.114	&	-0.243	\\
$^{199}$Hg	&	-0.167	&	-0.153	&	-0.014	&	-0.151	&	-0.016	\\
$^{201}$Hg	&	0.500	&	0.356	&	0.144	&	0.339	&	0.161	\\
$^{207}$Pb	&	-0.167	&	-0.162	&	-0.005	&	-0.161	&	-0.005	\\
  \end{tabular}%
    \label{tab:odd-neutron_isotopes}%
  \end{table}

Experimentally, the Schmidt model is known to overestimate the magnetic dipole moment in most nuclei. The simplest explanation for this is that the valence nucleon polarises the core nucleons, reducing the magnetic dipole moment of the nucleus. The degree of core polarisation can be estimated using experimental values of the magnetic dipole moment, and improved estimates for $\left<s_p^z\right>$ and $\left<s_n^z\right>$ can hence be obtained.

The reduction in nuclear magnetic dipole moment from the Schmidt value $\mu^0$ to the experimental value $\mu$ can proceed by a number of mechanisms. The simplest and most efficient way is to assume that the internucleon spin-spin interaction transfers spin from the valence proton (neutron) to core neutrons (protons):
\begin{equation}
\label{minimal}
\left(\left<s_p^z\right> - \left<s_p^z\right>^0\right) = - \left(\left<s_n^z\right> - \left<s_n^z\right>^0\right) = \frac{\mu - \mu^0}{g_p - g_n} ,
\end{equation}
where $\left<s_p^z\right>^0$ and $\left<s_n^z\right>^0$ are the Schmidt model values (one of which is necessarily zero). In general, there is also polarisation of the proton (neutron) core by the valence proton (neutron), but transfer of valence proton (neutron) spin to core proton (neutron) spin does not change the result. Note that the denominator $g_p - g_n = 9.412$ in (\ref{minimal}) is a large number, so the required change in $\left<s_p^z\right>$ and $\left<s_n^z\right>$ to obtain the experimental value $\mu$ is minimal. We present the values for $\left<s_p^z\right>$ and $\left<s_n^z\right>$ from Eq.~(\ref{minimal}) (``minimal model'') in Tables \ref{tab:odd-proton_isotopes} and \ref{tab:odd-neutron_isotopes}.

It is also possible for a reduction in nuclear magnetic dipole moment to occur by different mechanisms, for instance, by transfer of the spin angular momentum of a valence proton (neutron) to core proton (neutron) orbital angular momenta, or in a more unlikely manner by transfer of valence proton (neutron) spin angular momentum to core neutron (proton) orbital angular momenta. 

The preferred model of Refs.~\cite{Flambaum2006a,Berengut2011K} is intermediate to the two previously mentioned ``extreme models''. In this model, it is assumed that the total $z$ projections of proton and neutron angular momenta, $j_{p}^z$ and $j_{n}^z$, are separately conserved, and that the $z$ projections of total spin and orbital angular momenta, $\left<s_p^z\right> + \left<s_n^z\right>$ and $\left<l_p^z\right> + \left<l_n^z\right>$, are also separately conserved (which corresponds to the neglect of the spin-orbit interaction). In this case
\begin{eqnarray}
\label{s_pn}
\left<s_z\right>^0 = \left<s_p^z\right> + \left<s_n^z\right> , \\
\label{j_pz}
\left<j_{p}^z\right> = \left<s_p^z\right> + \left<l_p^z\right> ,
\end{eqnarray}
where $\left<j_{p}^z\right> = I$ for a valence proton and $\left<j_{p}^z\right> = 0$ for a valence neutron, with $\left<s_z\right>^0$ the Schmidt model value for the spin of the valence nucleon, given by (\ref{sz0}). From Eqs.~(\ref{mu-defn}), (\ref{s_pn}) and (\ref{j_pz}), we find
\begin{equation}
\label{s_n-pref}
\left<s_n^z\right> = \frac{\mu - \left<j_{p}^z\right> - (g_p - 1) \left<s_z\right>^0}{g_n - g_p +1} , \\
\end{equation}
\begin{equation}
\label{s_p-pref}
\left<s_p^z\right> = \left<s_z\right>^0 - \left<s_n^z\right> .
\end{equation}
We present the values for $\left<s_p^z\right>$ and $\left<s_n^z\right>$ from Eqs.~(\ref{s_n-pref}) and (\ref{s_p-pref}) (``preferred model'') in Tables \ref{tab:odd-proton_isotopes} and \ref{tab:odd-neutron_isotopes}. 

In the present work, we develop a new and alternate hybrid method, in which semi-empirical core-polarisation corrections are applied to \emph{ab initio} nuclear shell model calculations from Refs.~\cite{Engel1991,Ressell1997,Toivanen2009}. We use the results of the many-body calculations for $\mu^0$, $\left<s_p^z\right>^0$ and $\left<s_n^z\right>^0$ from Refs.~\cite{Engel1991,Ressell1997,Toivanen2009} as the input values (instead of the Schmidt model values) and improve them using the known experimental values of $\mu$. Minimal model corrections [from Eq.~(\ref{minimal})] to the proton and neutron spin angular momentum expectation values of the available nuclei are seen to generally reduce discrepancies in proton and neutron spin expectation values from different calculations, as shown in Table \ref{tab:hybrid}.

  \begin{table*}
    \centering%
    \caption{Expectation values $\left<s_n^z\right>$ and $\left<s_p^z\right>$ for selected nuclei after correcting \emph{ab initio} nuclear shell model spin expectation values via the minimal model correction scheme. For nuclei, where more than one calculation has been performed, we take the average of the final values of $\left<s_n^z\right>$ and $\left<s_p^z\right>$ for computing limits in the present work.} 
\begin{tabular}{ccccccc}
\multicolumn{1}{c}{Nucleus} & \multicolumn{1}{c}{Ref.} & \multicolumn{1}{c}{\emph{ab initio} model} & \multicolumn{1}{c}{$\left<s_n^z\right>^0$}   & \multicolumn{1}{c}{$\left<s_p^z\right>^0$}   & \multicolumn{1}{c}{$\left<s_n^z\right>$} & \multicolumn{1}{c}{$\left<s_p^z\right>$}   \\  
\hline
$^{125}$Te	&	\cite{Ressell1997}	&	Bonn A	&	0.287	&	0.001	&	0.274	&	0.014	\\
$^{125}$Te	&	\cite{Ressell1997}	&	Nijmegen II	&	0.323	&	-0.0003	&	0.297	&	0.026	\\
$^{127}$I	&	\cite{Ressell1997}	&	Bonn A	&	0.075	&	0.309	&	0.071	&	0.313	\\
$^{127}$I	&	\cite{Ressell1997}	&	Nijmegen II	&	0.064	&	0.354	&	0.100	&	0.318	\\
$^{127}$I	&	\cite{Toivanen2009}	&	Bonn-CD	&	0.030	&	0.418	&	0.108	&	0.340	\\
$^{129}$Xe	&	\cite{Ressell1997}	&	Bonn A	&	0.359	&	0.028	&	0.337	&	0.050	\\
$^{129}$Xe	&	\cite{Ressell1997}	&	Nijmegen II	&	0.300	&	0.0128	&	0.308	&	0.005	\\
$^{129}$Xe	&	\cite{Toivanen2009}	&	Bonn-CD	&	0.273	&	-0.0019	&	0.256	&	0.015	\\
$^{131}$Xe	&	\cite{Ressell1997}	&	Bonn A	&	-0.227	&	-0.009	&	-0.196	&	-0.040	\\
$^{131}$Xe	&	\cite{Ressell1997}	&	Nijmegen II	&	-0.217	&	-0.012	&	-0.187	&	-0.042	\\
$^{131}$Xe	&	\cite{Engel1991}	&	QTDA	&	-0.236	&	-0.041	&	-0.235	&	-0.042	\\
$^{131}$Xe	&	\cite{Toivanen2009}	&	Bonn-CD	&	-0.125	&	-0.00069	&	-0.122	&	-0.004	\\
$^{133}$Cs	&	\cite{Toivanen2009}	&	Bonn-CD	&	0.021	&	-0.318	&	-0.076	&	-0.221	\\
  \end{tabular}%
    \label{tab:hybrid}%
  \end{table*}

\section{Application I: Dark Matter Searches}
Proton and neutron spin contents are important for interpretations of experimental data from various dark matter detection schemes, which are based on effects involving couplings to nuclear spins. WIMP dark matter can undergo elastic, spin-dependent scattering off nuclei, see e.g.~\cite{Wilczek1985,DarkSide2012,Picasso2012,Zeplin2012,Fornasa2012,ArDM2013,EdelweissII2013,LUX2013,Cresst2014,DEAP-3600,Simple2014,SuperCDMS2014,Green2014}. Axions can induce oscillating nuclear Schiff moments via hadronic mechanisms \cite{Graham2011,Graham2013,Stadnik2014}, which can be sought for either directly through nuclear magnetic resonance-type experiments (CASPEr) \cite{Budker2013C} or oscillating atomic EDMs \cite{Stadnik2014}. Axions can interact directly with nuclear spins via the time-dependent spin-axion momentum coupling $\mathbf{s}_N \cdot \mathbf{p}_a \cos(m_a t)$, where $m_a$ is the axion mass \cite{Graham2013,Stadnik2014,Sikivie2014Atoms}, induce the time-dependent nuclear spin-gravity coupling $\mathbf{s}_N \cdot \mathbf{g} \cos(m_a t)$ and oscillating nuclear anapole moments \cite{Stadnik2014,RobertsCosmicPNC-long2014}. Magnetometry techniques can also be used to search for monopole-dipole and dipole-dipole axion exchange couplings \cite{Mainz2013exp,ArvanitakiGeraci2014}. Topological defect dark matter, which consists of axion-like pseudoscalar fields, can interact with nuclear spins via the time-dependent coupling $\mathbf{s}_N \cdot (\mathbf{\nabla} a)$, where $a$ is the pseudoscalar field comprising the topological defect \cite{Pospelov2013}, and can give rise to transient nuclear-sourced EDMs \cite{Stadnik2014defects}. Both of these effects can be sought for using GNOME \cite{Pustelny2013GNOME}. One may use Tables \ref{tab:odd-proton_isotopes}, \ref{tab:odd-neutron_isotopes} and \ref{tab:hybrid} for the interpretation of dark matter searches based on all of the mentioned schemes, as well as for tests of the fundamental symmetries of nature.

\section{Application II: Comagnetometer Experiments}
We first revisit the experiments of Refs.~\cite{Gemmel2010,Allmendinger2014}, in which a $^{3}$He/$^{129}$Xe comagnetometer was used to place constraints on the Standard Model Extension (SME) $\mathcal{CPT}$- and Lorentz-invariance-violating parameter $\tilde{b}_{\perp}^n$ \cite{Colladay1997-SME,Colladay1998-SME(LV)}, which quantifies the interaction strength of a background field with the spin of a neutron. The observed quantities are the amplitudes of sidereal frequency shifts, $\varepsilon_{1,X}$ and $\varepsilon_{1,Y}$, which in the case of the $^{3}$He/$^{129}$Xe system are related to the SME parameters via \cite{Kostelecky1999}:
\begin{equation}
\label{SME-Kost}
\left| 4\sin(\chi) \sum_{N=p,n} \left[\left<s_N^z\right>^{\textrm{(He)}} \tilde{b}_{J}^N - \frac{\gamma_{\textrm{He}}}{\gamma_{\textrm{Xe}}} \left<s_N^z\right>^{\textrm{(Xe)}} \tilde{b}_{J}^N \right]  \right| \leqslant 2\pi \varepsilon_{1,J}  ,
\end{equation}
where $J=X,Y$, $\gamma_{\textrm{He}}$ and $\gamma_{\textrm{Xe}}$ are the gyromagnetic ratios of $^{3}$He and $^{129}$Xe, respectively, with $\gamma_{\textrm{He}}/\gamma_{\textrm{Xe}} = 2.754$, and $\chi = 57^{\circ}$ is the angle between Earth's rotation axis and the quantisation axis of the spins. Within the Schmidt model, in which only valence neutrons participate in the spin-dependent coupling $\mathbf{s} \cdot \mathbf{\tilde{b}}$, it was determined that \cite{Allmendinger2014}:
\begin{eqnarray}
\label{Schmidt_result-simpleX}
\tilde{b}_{X}^n  = (4.1 \pm 4.7) \times 10^{-34} ~\textrm{GeV} , \\
\label{Schmidt_result-simpleY}
\tilde{b}_{Y}^n  = (2.9 \pm 6.2) \times 10^{-34} ~\textrm{GeV} .
\end{eqnarray}
However, in a non-single-particle model, proton spins also contribute. From our spin content values for $^{129}$Xe in Table \ref{tab:hybrid} and the values for the well-studied case of $^{3}$He from Ref.~\cite{Chupp1990}, we find, using Eq.~(\ref{SME-Kost}), instead of Eqs.~(\ref{Schmidt_result-simpleX}) and (\ref{Schmidt_result-simpleY}):
\begin{eqnarray}
\label{our_result-simpleX}
\tilde{b}_{X}^n + 0.20~ \tilde{b}_{X}^p = (9.2 \pm 10.5) \times 10^{-34} ~\textrm{GeV} , \\
\label{our_result-simpleY}
\tilde{b}_{Y}^n + 0.20~ \tilde{b}_{Y}^p = (6.5 \pm 13.9) \times 10^{-34} ~\textrm{GeV} ,
\end{eqnarray}
which gives the following limits ($1\sigma$) on $\tilde{b}_{\perp}^N = \sqrt{\left(\tilde{b}_{X}^N\right)^2+\left(\tilde{b}_{Y}^N\right)^2}$, where $N=p,n$, within the preferred model:
\begin{eqnarray}
\label{our_limit_bn}
|\tilde{b}_{\perp}^n| < 1.5 \times 10^{-33} ~\textrm{GeV} , \\
\label{our_limit_bp}
|\tilde{b}_{\perp}^p| < 7.6 \times 10^{-33} ~\textrm{GeV} .
\end{eqnarray}
Note that (\ref{our_limit_bp}) improves on the world's best proton-coupling limit of \cite{Brown2010} by a factor of 8 (Table \ref{tab:SME-sidereal_limits}). Thus in this case, the $^{3}$He/$^{129}$Xe system is sensitive not only to neutron SME parameters, but also has reasonable sensitivity to analogous proton parameters.

  \begin{table}[h!]
    \centering%
    \caption{Comparison of limits ($1\sigma$) on the SME parameters $\tilde{b}_{\perp}^n$ and $\tilde{b}_{\perp}^p$.} 
\begin{tabular}{cccc}
\multicolumn{1}{c}{Parameter} & \multicolumn{1}{c}{Ref.~\cite{Brown2010}}   &  \multicolumn{1}{c}{Ref.~\cite{Allmendinger2014}}  &  \multicolumn{1}{c}{This work}  \\ 
\hline
$|\tilde{b}_{\perp}^n|$ / GeV & $3.7 \times 10^{-33}$  & $8.4 \times 10^{-34}$  & $1.5 \times 10^{-33}$  \\  
$|\tilde{b}_{\perp}^p|$ / GeV & $6 \times 10^{-32}$  & --- & $7.6 \times 10^{-33}$  \\  
  \end{tabular}%
    \label{tab:SME-sidereal_limits}%
  \end{table}

Similarly, we reanalyse the results of Ref.~\cite{Cane2004}, in which a $^{3}$He/$^{129}$Xe comagnetometer was also used to place constraints on the SME parameters $\tilde{b}_{\perp}^n$, $\tilde{d}_{\perp}^n$ and $\tilde{g}_{ D\perp}^n$ \cite{Colladay1997-SME,Colladay1998-SME(LV)}, among others. The observed quantities are again the amplitudes of sidereal frequency shifts:
\begin{align}
\label{SME-Kost2}
\left| 4\sin(\chi) \sum_{N=p,n} \left\{ \left<s_N^z\right>^{\textrm{(He)}} \left[ \tilde{b}_{J}^N + \rho_N^{\textrm{(He)}} \tilde{d}_{J}^N + \tau_N^{\textrm{(He)}} \tilde{g}_{D J}^N \right] \right. \right. \\ \notag
\left. \left. -  \frac{\gamma_{\textrm{He}}}{\gamma_{\textrm{Xe}}} \left<s_N^z\right>^{\textrm{(Xe)}} \left[ \tilde{b}_{J}^N + \rho_N^{\textrm{(Xe)}} \tilde{d}_{J}^N + \tau_N^{\textrm{(Xe)}} \tilde{g}_{D J}^N \right] \right\} \right| \leqslant 2\pi \varepsilon_{1,J} ,
\end{align}
where
\begin{eqnarray}
\label{rho-N}
\rho_N = \left\{ \begin{array}{ll}
-\frac{\left< p^2 \right>_N}{(2l+3) m_N^2} & \textrm{if $j=l+\frac{1}{2}$,}\\
-\frac{ 3\left< p^2 \right>_N}{(2l+3) m_N^2} & \textrm{if $j=l-\frac{1}{2}$,}
\end{array} \right.
\end{eqnarray}
\begin{eqnarray}
\label{tau-N}
\tau_N = \left\{ \begin{array}{ll}
\frac{(l+1) \left< p^2 \right>_N}{(2l+3) m_N^2} & \textrm{if $j=l+\frac{1}{2}$,}\\
\frac{l \left< p^2 \right>_N}{(2l+3) m_N^2} & \textrm{if $j=l-\frac{1}{2}$.}
\end{array} \right.
\end{eqnarray}
Noting that the dominant contributions are from nucleons near the Fermi surface ($\approx 10$ MeV from the surface), taking the nucleon depth well to be $\approx 50$ MeV for both protons and neutrons, and using our spin content values for $^{129}$Xe in Table \ref{tab:hybrid} and the values for $^{3}$He from Ref.~\cite{Chupp1990}, along with the experimental data in Ref.~\cite{Cane2004}, we find the following results (all of which are consistent with zero):
\begin{align}
\label{new_result_X}
&\tilde{b}_{X}^n + 0.20~ \tilde{b}_{X}^p - 0.028 \tilde{d}_{X}^n - 0.006 \tilde{b}_{X}^p + 0.028 \tilde{g}_{DX}^n + 0.006 \tilde{g}_{DX}^p \\ \notag
&= (-5 \pm 18) \times 10^{-32}  ~\textrm{GeV} ,
\end{align}
\begin{align}
\label{new_result_Y}
&\tilde{b}_{Y}^n + 0.20~ \tilde{b}_{Y}^p - 0.028 \tilde{d}_{Y}^n - 0.006 \tilde{b}_{Y}^p + 0.028 \tilde{g}_{DY}^n + 0.006 \tilde{g}_{DY}^p \\ \notag
&= (1.8 \pm 2.1) \times 10^{-31}  ~\textrm{GeV} ,
\end{align}
where the uncertainties in the coefficients of $\tilde{d}_{J}^n$ and $\tilde{g}_{DJ}^n$ are a factor of several, while the uncertainties in the coefficients of $\tilde{d}_{J}^p$ and $\tilde{g}_{DJ}^p$ are an order of magnitude. We note that the corresponding sensitivities to the parameters $\tilde{d}_{\perp}^p$ and $\tilde{g}_{ D\perp}^p$ are at the level $\sim 10^{-29} - 10^{-28}$ GeV, which is a one order of magnitude improvement on the best corresponding proton-coupling sensitivities derived in \cite{Peck2012}.

Likewise, we revisit the experiment of Ref.~\cite{Tullney2013}, in which a $^{3}$He/$^{129}$Xe comagnetometer was used to place constraints on the spin-dependent $\mathcal{P}$,$\mathcal{T}$-violating interaction of a bound neutron with nucleons. The spin-dependent monopole-dipole coupling potential between two nucleons is given by \cite{Moody1984}:
\begin{equation}
\label{V_sp}
V_{sp}(\mathbf{r}) = \frac{g_s^N g_p^{N'}}{8 \pi m_{N'}} \left(\mathbf{\sigma} \cdot \mathbf{\hat{r}} \right) \left(\frac{1}{\lambda r} + \frac{1}{r^2} \right) e^{-r/\lambda} ,
\end{equation}
where $g_s^N$ is the dimensionless scalar coupling constant of the nucleon $N$ inside the spin-unpolarised sample, $g_p^{N'}$ is the dimensionless pseudoscalar coupling constant of the spin-polarised bound nucleon $N'$, $\mathbf{\hat{r}}$ is the unit vector from the bound nucleon to the unpolarised nucleon, $\mathbf{\sigma}$ is the spin of the polarised bound nucleon and $\lambda=1/m_{\textrm{a}}$ is the one-boson-exchange range. The resulting shift in the weighted frequency difference $\Delta \omega = \omega_{\textrm{He}} - \omega_{\textrm{Xe}} \gamma_{\textrm{He}}  / \gamma_{\textrm{Xe}}$ is given by (using results of derivations from Refs.~\cite{Tullney2013,Zimmer2010}):
\begin{equation}
\label{}
\Delta \nu_{sp}^{N'} = \left| 4 \sum_{N'=p,n} \left[ \left<s_N^z\right>^{\textrm{(He)}}  - \frac{\gamma_{\textrm{He}}}{\gamma_{\textrm{Xe}}} \left<s_N^z\right>^{\textrm{(Xe)}} \right] V_\Sigma^{N'} \right| ,
\end{equation}
with:
\begin{equation}
\label{}
V_\Sigma^{N'} = \frac{g_s^{N}g_p^{N'} N}{4 m_{N'}} \frac{\lambda^2}{D} e^{-\Delta x / \lambda} \left(1 - e^{-D/\lambda} \right) \left(1 - e^{-d/\lambda} \right) \eta(\lambda)  ,
\end{equation}
where $N$ is the number density of nucleons in the unpolarised sample, $D$ and $d$ are the thicknesses of the cylindrical spin-polarised and unpolarised samples, respectively, $\Delta x$ is the finite gap between the two samples and $\eta (\lambda)$ is a correction function accounting for the finite sizes of the two samples \cite{Tullney2013}.

\begin{figure}[h!]
\begin{center}
\includegraphics[width=8.5cm]{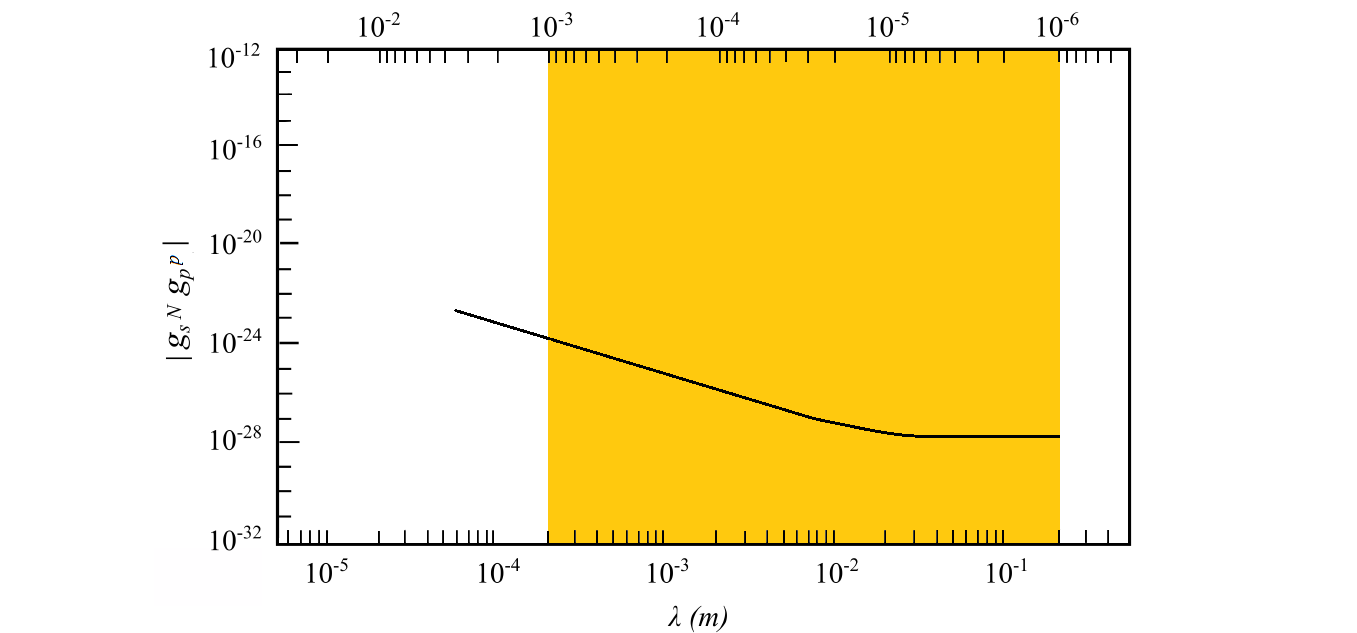}
\caption{$95\%$ confidence level upper limit on $\left| g_s^N g_p^p \right|$ as a function of the one-boson-exchange range $\lambda=1/m_{\textrm{a}}$. Solid black line corresponds to limits derived in our present work. Shaded orange region indicates `classical' region of axion masses.} 
\label{fig:proton_limits}
\end{center}
\end{figure}

\begin{figure}[h!]
\begin{center}
\includegraphics[width=8.5cm]{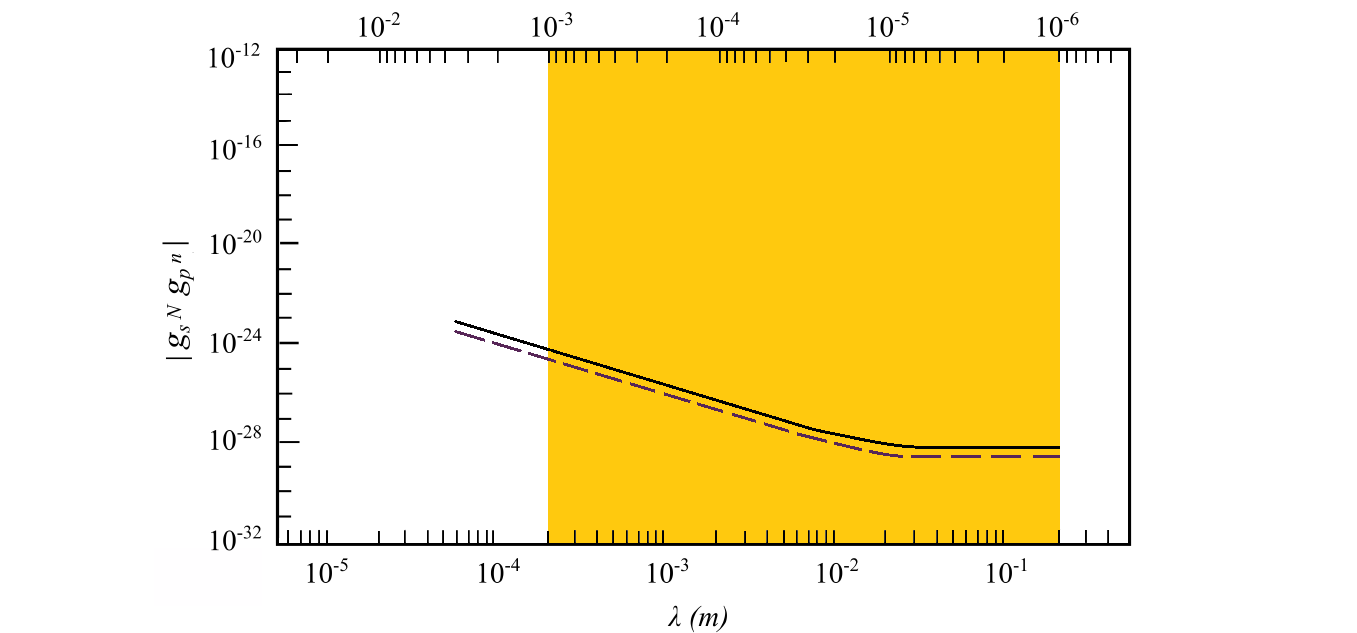}
\caption{$95\%$ confidence level upper limit on $\left| g_s^N g_p^n \right|$ as a function of the one-boson-exchange range $\lambda=1/m_{\textrm{a}}$. Solid black line corresponds to limits derived in our present work. Dashed purple line corresponds to limits obtained from Schmidt model in Ref.~\cite{Tullney2013}. Shaded orange region indicates `classical' region of axion masses.}
\label{fig:neutron_limits}
\end{center}
\end{figure}

Combining the experimental data of \cite{Tullney2013} with our spin content values for $^{129}$Xe in Table \ref{tab:hybrid} and the values for $^{3}$He from Ref.~\cite{Chupp1990}, we obtain the 95$\%$ confidence level upper limits on the parameters $\left| g_s^N g_p^p \right|$ and $\left| g_s^N g_p^n \right|$ shown in Figures \ref{fig:proton_limits} and \ref{fig:neutron_limits}, respectively. For some of the other limits on these parameters, we refer the reader to Refs.~\cite{Kimball2014_Nuc,Youdin1996,Baesler2007,Glenday2008,Serebrov2010,Petukhov2010,Hoedel2011,Raffelt2012,Jenke2012,Bulatowics2013,Chu2013}. 

\section{Application III: Tests of Fundamental Symmetries}
Consider the following Lorentz-invariance-violating terms in the SME Lagrangian (in the natural units $\hbar = c =1$) \cite{Kostelecky1999}:
\begin{equation}
\label{SME_L}
\mathcal{L} = -b_{\mu} \bar{\psi} \gamma_5 \gamma^\mu \psi + \frac{i}{2} d_{\mu \nu} \bar{\psi} \gamma_5 \gamma^\mu \stackrel{\leftrightarrow}{\partial^\nu} \psi ,
\end{equation}
where $b_{\mu}$ and $d_{\mu \nu}$ are background fields, $\psi$ is the fermion wavefunction with $\bar{\psi} \equiv \psi^\dagger \gamma^0$, $\gamma^0$, $\gamma_5$ and $\gamma^\mu$ are Dirac matrices, and the two-sided derivative operator $\stackrel{\leftrightarrow}{\partial^\nu}$ is defined by: $A\stackrel{\leftrightarrow}{\partial^\nu} B \equiv A (\partial^\nu B) - (\partial^\nu A) B$. The first term in (\ref{SME_L}) is $\mathcal{CPT}$-odd, while the second term is $\mathcal{CPT}$-even. In the non-relativistic limit, the Lagrangian (\ref{SME_L}) gives rise to the following interaction Hamiltonian
\begin{equation}
\label{SME_H_NRL}
H_{\textrm{int}} =  \frac{2 b_0}{m} \mathbf{s} \cdot \mathbf{p} - 2d_{00} \mathbf{s} \cdot \mathbf{p} ,
\end{equation}
where $m$ is the fermion mass, $\mathbf{s}$ is the fermion spin and $\mathbf{p}$ is the fermion momentum operator. In our previous work \cite{Roberts2014}, we showed in the single-particle approximation that the first term in (\ref{SME_H_NRL}) gives rise to nuclear anapole moments associated with valence nucleons \cite{Flambaum1984} (see also \cite{Stadnik2014}). Experimentally, the nuclear anapole moment manifests itself as a NSD contribution to a PNC amplitude. Hence from the measured and calculated (within the SM) values of the anapole moments of Cs and Tl, we were able to extract direct limits on the parameter $b_0^p$.

In the single-particle approximation, the nuclear anapole moment contribution from interaction (\ref{SME_H_NRL}) is
\begin{equation}
\label{AM_CF}
\mathbf{\tilde{a}}^N = \frac{G_F}{\sqrt{2}e} \frac{K\mathbf{I}}{I(I+1)} (\kappa_b^N + \kappa_d^N) ,
\end{equation}
where $G_F$ is the Fermi constant of the weak interaction, $K=(I+1/2)(-1)^{I+1/2-l}$, and the dimensionless constants $\kappa_b^N$ and $\kappa_d^N$ are given by
\begin{eqnarray}
\label{kappa_b}
\kappa_b^N = \frac{2\sqrt{2}\hbar \pi \alpha \mu_N \left<r^2\right> b_0^N}{G_F m_N c} , \\
\label{kappa_d}
\kappa_d^N = - \frac{2\sqrt{2}\hbar \pi \alpha \mu_N \left<r^2\right> d_{00}^N}{G_F c} ,
\end{eqnarray}
where $\alpha = e^2/\hbar c$ is the fine-structure constant, $m_N$ and $\mu_N$ are the mass and magnetic dipole moment of the unpaired nucleon $N$ ($\mu_p=2.8$ and $\mu_n=-1.9$), respectively, and we take the mean-square radius $\left<r^2\right> = \frac{3}{5}r_0^2 A^{2/3}$, with $r_0 = 1.2$ fm, and $A$ the atomic mass number. Combining the measured values for the nuclear anapole moment of $\kappa_a{\rm(Cs)}=0.364(62)$~\cite{Wood1997_Cs-PNC,Flambaum1997-AM-Murray} and $\kappa_a{\rm(Tl)}=-0.22(30)$~\cite{Vetter1995_Tl-PNC,Khriplovich1995}, with the values $\kappa_a{\rm(Cs)}=0.19(6)$ and $\kappa_a{\rm(Tl)}=0.17(10)$ from nuclear theory~\cite{Dmitriev1997,Dmitriev2000,Haxton2001c,Haxton2002} (see also~\cite{Ginges_PhysRep2004}), and with Eq.~(\ref{kappa_d}), we extract limits on the parameter $d_{00}^p$ in the single-particle approximation (Table \ref{tab:SME-anapole_moments}).

  \begin{table}[h!]
    \centering%
    \caption{New limits ($1\sigma$, in laboratory frame) on the SME parameters $b_0^p$, $d_{00}^p$, $b_0^n$ and $d_{00}^n$. s.p.~denotes single-particle (Schmidt model) limit and m.b.~denotes many-body (hybrid model) limit.} 
\begin{tabular}{cccccc}
 & & \multicolumn{2}{c}{Ref.~\cite{Roberts2014}}     &  \multicolumn{2}{c}{This work}  \\ 
\cline{3-4} \cline{5-6} 
\multicolumn{1}{c}{Parameter} & \multicolumn{1}{c}{Model} & Cs & Tl & Cs & Tl \\
\hline
$|b_{0}^p|$ / GeV & s.p. & $3 \times 10^{-8}$  & $8 \times 10^{-8}$  & --- & ---  \\  
$|d_{00}^p|$ 	& s.p.   & --- & --- & $3 \times 10^{-8}$  & $9 \times 10^{-8}$  \\  
$|b_{0}^p|$ / GeV & m.b.   & --- & ---  & $7 \times 10^{-8}$  & --- \\  
$|d_{00}^p|$  & m.b.   & --- & ---  & $8 \times 10^{-8}$  & --- \\  
$|b_{0}^n|$ / GeV & m.b.   & --- & ---  & $3 \times 10^{-7}$  & --- \\  
$|d_{00}^n|$  & m.b.   & --- & ---  & $3 \times 10^{-7}$  & --- \\  
  \end{tabular}%
    \label{tab:SME-anapole_moments}%
  \end{table}

We now leave the single-particle approximation and consider nuclear many-body effects. For a single-particle state, the angular momenta factors in (\ref{AM_CF}) can be rewritten as 
\begin{eqnarray}
\label{AMo-factors}
\frac{K\mathbf{I}}{I(I+1)} = \left\{ \begin{array}{ll}
-\frac{2(I+1/2)}{I+1} \left< \mathbf{s} \right> & \textrm{if $I=l+\frac{1}{2}$,}\\
-\frac{2(I+1/2)}{I} \left< \mathbf{s} \right> & \textrm{if $I=l-\frac{1}{2}$.}
\end{array} \right.
\end{eqnarray}
Hence, unlike NSD-PNC effects arising from $Z^0$-boson exchange between electrons and the nucleus \cite{Ginges_PhysRep2004}, we cannot simply average over the spins of the single-particle proton and neutron states without explicitly considering the angular momenta of each individual nucleon. To circumvent this difficulty, we make use of the following approximation. Note that for single-particle states with $j>1$, the prefactors before $\left< \mathbf{s} \right>$ in Eq.~(\ref{AMo-factors}) are $\approx -2$. For non-light nuclei, most nucleons have $j>1$. Also, the deviations of the prefactors in (\ref{AMo-factors}) from $-2$ are of opposite sign for $j=l \pm 1/2$. Thus for nuclei with valence nucleon(s), which have $j >1$, we can approximately sum over the proton and neutron spin angular momenta that appear in (\ref{AMo-factors}) to give the many-body generalisation of formula (\ref{AM_CF}):
\begin{equation}
\label{AM_MB-gen}
\mathbf{\tilde{a}} \approx \frac{-\sqrt{2} G_F}{e} \left[ (\kappa_b^p + \kappa_d^p) \left< \mathbf{s}_p \right> + (\kappa_b^n + \kappa_d^n) \left< \mathbf{s}_n \right> \right] . 
\end{equation}
From Eq.~(\ref{AM_MB-gen}), we extract limits on the parameters $b_{0}^p$, $d_{00}^p$, $b_{0}^n$ and $d_{00}^n$ for Cs, for which $I=7/2$, using the calculated spin content values in Table \ref{tab:hybrid}. The limits are presented in Table \ref{tab:SME-anapole_moments}. For Tl, where $I=1/2$, Eq.~(\ref{AM_MB-gen}) is not a good approximation and so we do not present many-body model limits in this case. We note that the limits in Table \ref{tab:SME-anapole_moments} are weaker than those that would be obtained indirectly from the most stringent limits on $\tilde{b}^p_i$ and $\tilde{b}^n_i$, if one assumes a static background cosmic field. These corresponding upper limits are roughly as follows: $\left|b_0^n \right| \lesssim 10^{-29}$ GeV, $\left|d_{00}^n \right| \lesssim 10^{-29}$, $\left|b_0^p \right| \lesssim 10^{-28}$ GeV and $\left|d_{00}^p \right| \lesssim 10^{-28}$, assuming that the typical speed of Earth relative to the static background cosmic field is $v \sim 10^{-4} - 10^{-3} c$.

\begin{acknowledgements}
We would like to thank Dmitry Budker and Lutz Trahms for useful discussions and for motivating this work. We are particularly grateful to Alan Kosteleck\'y for pointing out that the methods in our previous work \cite{Roberts2014} could be extended to extract limits on the SME parameter $d_{00}^p$ and that the methods of our present work could be extended to extract sensitivities on the SME parameters $\tilde{d}_{\perp}^p$ and $\tilde{g}_{ D\perp}^p$. Y.~V.~S.~would like to thank Pierre Sikivie for useful discussions and for suggesting several further nuclei of interest in axion dark matter searches. The authors would also like to thank Ben Roberts for useful discussions. This work was supported in part by the Australian Research Council and by the Perimeter Institute for Theoretical Physics. Research at the Perimeter Institute is supported by the Government of Canada through Industry Canada and by the Province of Ontario through the Ministry of Economic Development \& Innovation. V.~V.~F.~would also like to acknowledge the Humboldt foundation for support through the Humboldt Research Award and the MBN Research Center for hospitality. 
\end{acknowledgements}





\end{document}